\documentclass[sigconf, authorversion,nonacm]{acmart}

\AtBeginDocument{%
  \providecommand\BibTeX{{%
    \normalfont B\kern-0.5em{\scshape i\kern-0.25em b}\kern-0.8em\TeX}}}


\setcopyright{none}
\renewcommand\footnotetextcopyrightpermission[1]{} 


\copyrightyear{2023}
\acmYear{2023}
\setcopyright{rightsretained}
\acmConference[SIGCSE 2023]{Proceedings of the 54th ACM Technical Symposium on Computing Science Education V. 2}{March 15--18, 2023}{Toronto, ON, Canada}
   
\acmBooktitle{Proceedings of the 54th ACM Technical Symposium on Computer Science Education V. 2 (SIGCSE 2023), March 15--18, 2023, Toronto, ON, Canada}
\acmDOI{10.1145/3545947.3573358} \acmISBN{978-1-4503-9433-8/23/03}

\begin{document}

\title{Rethinking Computing Education Assessment after Generative AI}
\title{Reimagining Computing Education Assessment after Generative AI}
\title{Imagining Computing Education Assessment after Generative AI}



\author{Stephen MacNeil}
\affiliation{%
  \institution{Temple University}
  \city{Philadelphia}
  \state{PA}
  \country{USA}
}
\email{stephen.macneil@temple.edu}
\orcid{0000-0003-2781-6619}

\author{Scott Spurlock}
\affiliation{%
  \institution{Elon University}
  \city{Elon}
  \state{NC}
  \country{USA}}
\email{sspurlock@elon.edu}
\orcid{0000-0002-5090-7837}

\author{Ian Applebaum} 
\affiliation{%
  \institution{Temple University}
  \city{Philadelphia}
  \state{PA}
  \country{USA}
}
\email{ian.tyler@temple.edu}
\orcid{0009-0002-0704-9379}




\renewcommand{\shortauthors}{MacNeil, et al.}

\begin{abstract}

In the contemporary landscape of computing education, the ubiquity of Generative Artificial Intelligence has significantly disrupted traditional assessment methods, rendering them obsolete and prompting educators to seek innovative alternatives. This research paper explores the challenges posed by Generative AI in the assessment domain and the persistent attempts to circumvent its impact. Despite various efforts to devise workarounds, the academic community is yet to find a comprehensive solution. Amidst this struggle, ungrading emerges as a potential yet under-appreciated solution to the assessment dilemma. Ungrading, a pedagogical approach that involves moving away from traditional grading systems, has faced resistance due to its perceived complexity and the reluctance of educators to depart from conventional assessment practices. However, as the inadequacies of current assessment methods become increasingly evident in the face of Generative AI, the time is ripe to reconsider and embrace ungrading.

\end{abstract}





\maketitle

\section{Introduction}


Large language models are a powerful new tool that students and professionals are using to support their learning and to facilitate their work. These models have been shown capable of writing high-quality code~\cite{chen2021evaluating, barke2022grounded,sobania21arxiv, denny2023conversing}, identify coding concepts~\cite{tran2023using}, answering multiple-choice questions~\cite{savelka2023generative, savelka2023thrilled}, and even solving parsons problems~\cite{reeves2023evaluating}. These use cases are exciting for professionals, but researchers and educators are concerned about their potential misuse and challenges related to academic integrity~\cite{prather2023WG-abstract, becker2023programming, zastudil2023generative, lau2023ban}. For example, Lau et al.'s interview study of 20 globally distributed educators surfaced short-term beliefs that it should be banned with longer-term beliefs that `resistance is futile.'




\subsection{Generative AI Threats for Assessment}

Educators are currently very concerned about the impacts that generative AI tools will have on assessment~\cite{prather2023WG-abstract, becker2023programming, zastudil2023generative, lau2023ban}. Generative AI tools can write code for students~\cite{barke2022grounded,sobania21arxiv, denny2023conversing}, identify bugs~\cite{macneil2024decoding}, and perform well on quizzes~\cite{savelka2023thrilled, reeves2023evaluating}, including visual programming problems~\cite{hou2024more}. Solutions are beginning to emerge, such as heavily-weighted proctored exams; however, these solutions require significant effort from educators and are misaligned with students preferences~\cite{zastudil2023generative} and conflict with trends toward lightweight assessment methods designed to reduce test anxiety~\cite{latulipe2018evolving, macneil2016exploring, latulipe2015structuring}.




\section{Forging a Path Forward}

While some educators lament the loss of traditional assessments, our group welcomes it as a chance for positive change. Often, assessments have been unfair or wielded as a means of coercion to ensure students `do the work.' The proliferation of generative AI tools eliminates this enforcement mechanism, pushing educators to adopt student-centered teaching methods. \textit{With grades no longer driving student motivation, we need to reevaluate how to inspire students effectively.} By drawing on concepts like ungrading and active learning, we explore new pathways to enrich the learning experience and nurture intrinsic motivation in students.





\subsection{Reimagining Assessment} 

Assessment has historically served three main purposes: feedback, credentialing, and motivation. First and foremost, assessment is an opportunity for feedback. Through assessment, students understand how they are doing and where they can improve. Second, assessment supports credentialing, i.e., a signal that differentiates students for hiring purposes. And third, assessment provides student motivation in the form of grades. While the role of assessment in providing formative feedback is key to student learning, there may be reason to doubt its effectiveness for credentialing and motivation. In fact, research stretching back nearly a century has shown that traditional numeric grades are not a particularly reliable signal of student achievement, casting doubt on their suitability for identifying qualified job applicants~\cite{brookhart2016century,crooks1933marks,meadows2005review,kohn1999degrading}. Further, traditional grading appears to exacerbate equity concerns among traditionally underrepresented groups of students~\cite{schinske2014teaching,feldman2018grading}. Also dubious is the use of grades for student motivation. Grades serve as an \textit{extrinsic motivator}, while research suggests that \textit{intrinsic motivators} are better associated with student learning~\cite{kohn1999degrading}. Several studies have found improved learning and attitude among students receiving feedback rather numeric grades~\cite{butler1988enhancing,spurlock2023improving}.

There is also a correlation between traditional assessment and student cheating. While increased intrinsic motivation can reduce cheating~\cite{murdock2006motivational}, pressure, opportunity, and rationalization are associated with increased incidence~\cite{albluwi2019plagiarism}. Although most instructor effort typically goes into reducing opportunity by making cheating difficult through assignment design or incorporating tools that detect plagiarism, less focus has been given to enhancing the learning environment to reduce the motivation to cheat~\cite{albluwi2019plagiarism}. Indeed, efforts to improve student support and inclusivity may be more effective~\cite{wortzman2022s}.

Based on these and other issues with traditional forms of grading, there has been a recent surge of interest in alternative forms of assessment, typified by Blum's \textit{Ungrading}~\cite{blum2020ungrading}, which seeks to deemphasize the role of numeric scores in favor of other forms of feedback. Ungrading approaches advocate non-traditional ideas, such as allowing students to play a role in their own assessment~\cite{chu2020} and encouraging multiple submissions of assignments~\cite{riesbeck2020}. While these ideas are undergoing something of a renaissance, approaches such as mastery grading, which advocates for repeated attempts until a student has achieved mastery of a topic, have been around for decades~\cite{bloom1968learning}. Recent work has found a link between ungrading and improving intrinsic motivation~\cite{spurlock2023improving}, suggesting that traditional assessment methods may not truly be needed to motivate students. Other novel pedagogical approaches have also shown success in sparking student motivation. 








It is clear that the reduced emphasis on traditional assessment will necessitate a reconsideration of incentives for students to master course material.
Adopting new pedagogical practices demands substantial effort, including the willingness to be vulnerable and to experience encountering repeated failures. Yet, adapting our teaching and assessment methodologies is no longer merely an option; it is imperative in light of tools that fundamentally reshape how students learn and the nature of their work. One compelling avenue is a heightened emphasis on intrinsic motivation.



\section{Conclusion} 

In this short position piece, we hope to redefine the discourse surrounding assessment by shifting the focus from reviving traditional evaluation methods to embracing the historical and contemporary trends in ungrading and intrinsic motivation. Our contribution lies in three key aspects: firstly, outlining the challenges and threats confronting traditional assessment; secondly, summarizing the historical trends and current state of ungrading; and thirdly, reframing the current assessment crisis as a matter of student motivation. By doing so, the poster charts a course that prioritizes learners, redirecting attention from external motivators to intrinsic sources of inspiration. Positioned as a catalyst for dialogue and innovation, the overarching aim of this poster is to foster a dynamic community of practice. This community includes educators, researchers, and stakeholders deeply engaged in the transformative intersection of generative AI and assessment methodologies.

\bibliographystyle{ACM-Reference-Format}
\bibliography{sample-base}

\end{document}